\documentclass{elsart1p}

\usepackage{amssymb}
\usepackage{epsfig}
\usepackage{graphics,color}
\newcommand{\be}{\begin{eqnarray}}
\newcommand{\ee}{\end{eqnarray}}
\newcommand{\vck}{{\bf k}}
\newcommand{\vcx}{{\bf x}}

\begin{document}

\begin{frontmatter}

% Title, authors and addresses

% use the thanksref command within \title, \author or \address for footnotes;
% use the corauthref command within \author for corresponding author footnotes;
% use the ead command for the email address,
% and the form \ead[url] for the home page:
% \title{Title\thanksref{label1}}
% \thanks[label1]{}
% \author{Name\corauthref{cor1}\thanksref{label2}}
% \ead{email address}
% \ead[url]{home page}
% \thanks[label2]{}
% \corauth[cor1]{}
% \address{Address\thanksref{label3}}
% \thanks[label3]{}

\title{2PI simulations in expanding backgrounds:\\ Doing it anyway!}

% use optional labels to link authors explicitly to addresses:
% \author[label1,label2]{}
% \address[label1]{}
% \address[label2]{}

\author{Anders Tranberg}
\ead{anders.tranberg@oulu.fi}

%\address{\thanks{email: anders.tranberg@oulu.fi}\\
% {\em\normalsize Department of Physical Sciences, University of Oulu} \\
%   {\em\normalsize P.O. Box 3000, FI-90014 Oulu, Finland}}
\address{Department of Physical Sciences, University of Oulu\\ P.O. Box 3000, FI-90014 Oulu, Finland}

\begin{abstract}
I set out the 2PI formalism for quantum fields in a Friedmann-Robertson-Walker Universe. I show how one can solve self-consistently for the quantum field evolution and the evolution of the cosmological scale factor, using a renormalised semi-classical Friedmann equation. In a 2PI coupling expansion truncated at $\mathcal{O}(\lambda^2)$, I perform sample calculations demonstrating kinetic and chemical equilibration in this context. 
\end{abstract}

%\begin{keyword}
% keywords here, in the form: keyword \sep keyword

% PACS codes here, in the form: \PACS code \sep code
%\PACS 
%\end{keyword}
\end{frontmatter}

% main text

% SECTION: INTRODUCTION

\section{Introduction and setup}
Direct numerical calculations in quantum field theory often rely on lattice discretization of certain equations of motion, which are then solved in real time. Although this is straightforward in Minkowski space, where the lattice is static, when transferred to an expanding Universe, a number of issues need to be resolved. Firstly, the scale factor of the Friedmann-Robertson-Walker Universe now multiplies the lattice spacing (the lattice coordinates are comoving), which means that discretization errors increase in time (the available momentum modes {\it redshift out of the box}). Given a finite lattice to start with, this imposes a cap on the amount of expansion and hence the time scales one can hope to reliably simulate. Secondly, the dynamics of the scale actor is determined by the energy momentum tensor of matter through the Friedmann equation, a quantity which in a quantum theory must be renormalised.  

The 2PI formalism for quantum fields in real time has been very successful recently in describing the approach to equilibrium and thermalisation (see for instance \cite{Berges:2000ur,Arrizabalaga:2005tf}). It is ideally suited to study strongly out of equilibrium phenomena like post-inflationary (p)reheating and phase transitions in the early Universe. However, unless we can argue that expansion is negligible (and sometimes we can), we need to extend the formalism to include this expansion. 

The 2PI equations of motion turn out to generalise very simply, and although an exact renormalisation is cumbersome, for practical numerical purposes there is a well defined and systematically improvable procedure by which one can renormalise at finite lattice spacing to the required accuracy. 

At the end of day the issue of redshift is a question of having a large enough lattice on a large enough computer, and can not be avoided. The approach here is to instead bypass the problem by only considering the phenomena, for which the expansion range fits on available lattice sizes. A back-of-the envelope estimate yields that $a(t_{stop})/a(t_{start})<n_x/2\pi$, where $a(t)$ is the scale factor, and $n_x$ is the linear scale of the lattice.
This essentially rules out direct studies of inflation, but since $T\propto a^{-1}$, it allows a large range in temperatures and therefore most post-inflationary phenomena.
%
%
% SECTION: EQUATIONS OF MOTION
%
%\subsection{Equations of motion}
%
We will study a single self-interacting scalar field in an FRW Universe with scale factor $a(t)$, with the action
\begin{equation}
S= \int dt\,d^{3}\vcx\, a^{3}(t)\bigg[ \frac{1}{2}(\partial_t\varphi)^{2}-\frac{1}{2a^2(t)}(\partial_{\bf x}\varphi)^{2}-\frac{1}{2}m^2\varphi^2-\frac{\lambda}{24}\varphi^4\bigg].
\end{equation}
The scale factor is treated as an external field, and evolves according to the semi-classical Friedmann equation $3M_{\rm pl}^2H^2(t)=\langle T^{00}(t)\rangle_{\rm ren}$, where $H(t)=\frac{\dot{a}(t)}{a(t)}$
is the Hubble rate, and the energy momentum $T^{00}$ is renormalised relative to the adiabatic vacuum (see below). The 2PI equations of motion can then be derived in the usual way. Going to conformal time $\eta$, $dt=a(\eta)d\eta$, and rescaling the field $\phi({\bf x},\tau)/a(\tau)=\varphi({\bf x},t)$, we have
\be
\left[\partial_\eta^2+k^2+M^2(\eta)\right]F(\eta,\eta',{\bf k})&=&-\int_{0}^{\eta}d\eta'' \Sigma_\rho(\eta,\eta'',{\bf k})F(\eta'',\eta',{\bf k})\nonumber\\
&&+\int_{0}^{\eta'}d\eta'' \Sigma_F(\eta,\eta'',{\bf k})\rho(\eta'',\eta',{\bf k}),\\
\left[\partial_\eta^2+k^2+M^2(\eta)\right]\rho(\eta,\eta',{\bf k})&=&-\int_{\eta'}^{\eta}d\eta'' \Sigma_\rho(\eta,\eta'',{\bf k})\rho(\eta'',\eta',{\bf k}),\\
\left[\partial_\eta^2+M^2_\phi(\eta)\right]\bar\phi(\eta)&=&-\int_0^\eta d\eta' \int d^3{\bf x}\,\Sigma_\phi(\eta,\eta',{\bf x})\bar\phi(\eta'),
\ee
in terms of the mean field $\bar\phi(\eta)=\langle\phi(x)\rangle$ and the two-point function $\langle T\phi(x)\phi(y)\rangle-\bar\phi(\eta)\bar\phi(\eta')=F(\eta,\eta',{\bf x-y})-\frac{i}{2}\rho(\eta, \eta',{\bf x-y}){\rm \,sign}_{\mathcal{C}}(\eta-\eta')$, . The precise form of the self-energy components $M^2_\phi$, $M^2$, $\Sigma_\phi$, $\Sigma_F$, $\Sigma_\rho$ depends on the truncation of a diagram expansion, in this case a coupling expansion to $\mathcal{O}(\lambda^2)$, and can be found in \cite{Berges:2000ur,Arrizabalaga:2005tf,Tranberg:2008ae}. In the rescaled field, conformal time variables, the expansion of the Universe only enters through an effective modification of the masses 
$m^2\rightarrow-\frac{a''(\eta)}{a(\eta)}+a^2(\eta)m^2$
where $a''=d^2a/d\eta^2$, the trivial rescaling of the momenta $k_{\rm phys}=k/a(\eta)$, and the fact that time is stretched through $t-t_0=\int^\eta_{\eta_0} a(\eta) d\eta$ (see also \cite{Calzetta} for related or partial implementations of expanding backgrounds). 
\begin{figure}
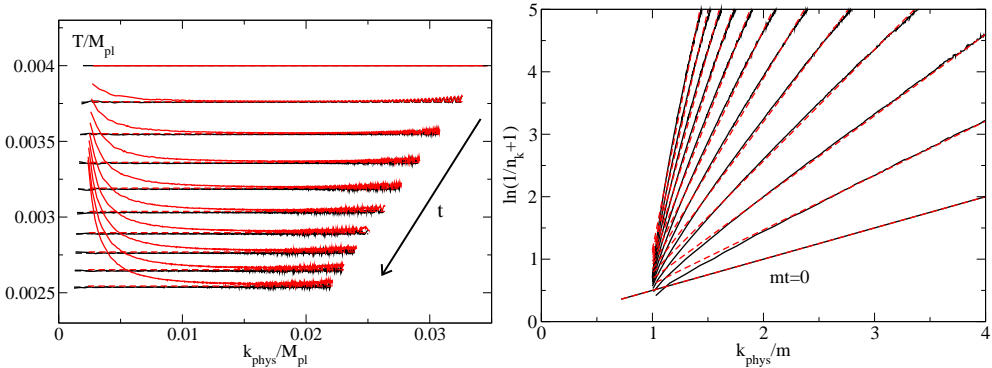

\begin{center}
\epsfig{file=./modes_T2.eps,width=6.5cm,clip}
\epsfig{file=./spectrum_T2int_ex.eps,width=6.5cm,clip}
\caption{Left: The effective temperature $T=\omega_\vck/\ln(1/n_\vck+1)$ for each momentum mode $k$ at different times, assuming a Bose-Einstein spectrum $n_\vck$. Shown are the cases of a free massless (black) and massive (red) field. The massive case is driven away from equilibrium, and an effective chemical potential is generated. Time increases downwards, and the physical momentum range shrinks due to redshift. Right: $\omega_\vck/T$ vs. $k$ for an interacting massive field (black) compared to Bose-Einstein fits (red), which allow us to extract effective temperatures and chemical potentials.}
\label{fig:T2modes}
\end{center}
\end{figure}
\begin{figure}
\begin{center}
\epsfig{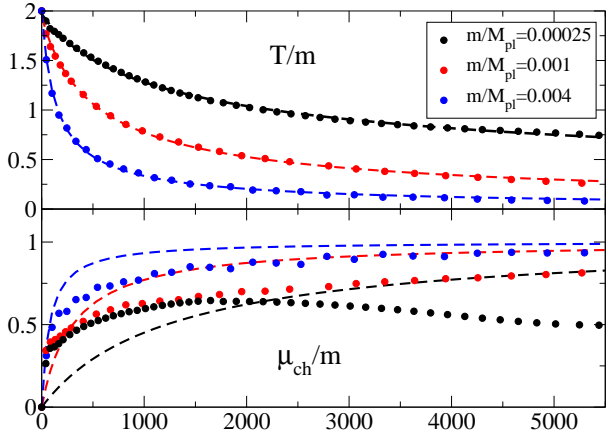}
\caption{Effective temperatures (upper) and chemical potentials (lower) in time for an interacting massive field, initially in equilibrium, for different expansion rates. Blue is fastest, black slowest. Temperature decreases as expected, but the chemical potentials increase due to the expansion as for a free field (dashed lines). Only in the slowest expansion case do the interactions manage to partially compensate for the expansion, chemically equilibrating back towards zero chemical potential.}
\label{fig:T2modes2}
\end{center}
\end{figure}
%
% SECTION: RENORMALISATION
%
%\subsection{Renormalisation\label{renorm}}
%
Renormalisation of the energy momentum tensor is ambiguous, and requires specification of the vacuum with respect to which we renormalise. Here we choose the vacuum defined by solving the vacuum 2PI-LO (Hartree) equations as a function of $a(t)$. In practice we solve them to second order in a WKB approximation and to second order in derivatives of $a(t)$ (order $H^2$, $\dot{H}$). We can then define a mass and an energy counterterm, whereby we cancel quartic and quadratic divergences, while neglecting logarithmic ones, at 2PI-LO. By going to higher order in WKB and $H$ one can take care of the log divergences as well, but this is beyond our scope. Also, diagrams at NLO ($\lambda^2$) introduce new divergences of all orders, but these are suppressed by the coupling. Details of this can be found in \cite{Anderson:2005hi,Tranberg:2008ae}. In practice, the energy momentum is renormalised to zero in the vacuum with a relative accuracy of $10^{-5}$ for the simulation parameters used below.
\begin{figure}[t!]
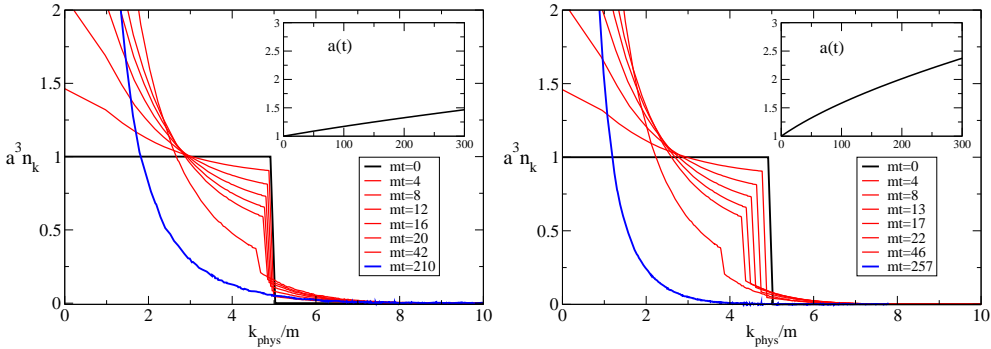

\begin{center}
\epsfig{file=./tsuspectrum200_1.eps,width=6.5cm,clip}
\epsfig{file=./tsuspectrum50_1.eps,width=6.5cm,clip}
\caption{The particle number $n_\vck$ vs. momentum for different time, starting from a way-out-of equilibrium state (black). Both for slow (left) and fast (right) expansion, interactions are strong enough to kinetically equilibrate to a final Bose-Einstein distribution (blue line). Again, we can extract effective temperatures and chemical potentials, to find that only at the slowest expansion rates does chemical equilibration take place. With fast expansion, a finite chemical potential persists.}
\label{fig:T2modes3}
\end{center}
\end{figure}
%
% SECTION: SAMPLE APPLICATIONS
%
\section{Sample applications}
We performed sample simulations using $n_x^3=32^3$ lattices starting at $a(0)m=0$ (massless) and $a(0)m=0.2$ (massive), expanding by a factor of order $10$. The coupling was $\lambda=6$, for which the coupling expansion is known to converge, and the LO renormalisation approximation is reasonable. The expansion rate is encoded through the Planck mass $M_{\rm pl}$, with larger Planck mass leading to slower expansion. Applications include: Free field originally in equilibrium in an expanding background (Fig. \ref{fig:T2modes}, left). The same for an interacting field (Fig. \ref{fig:T2modes}, right and \ref{fig:T2modes2}). An interacting field initially far from equilibrium (Fig. \ref{fig:T2modes3}). More applications can be found in \cite{Tranberg:2008ae}, including an oscillating free mean field (inflaton) and resonant preheating.

\vspace{0.3cm}

\noindent{\bf Acknowledgments:}
The numerical work was conducted on the Murska cluster at the Finnish center for computational sciences, CSC. This work was supported by Academy of Finland Grant 114371.
%
% THE BIBLIOGRAPHY
%

\end{document}